\documentclass[12pt]{article}
\begin{document}

\begin{center}
{\large \bf Logic and numbers related to solar neutrinos}
\vspace{0.5 cm}

\begin{small}
\renewcommand{\thefootnote}{*}
L.M.Slad\footnote{slad@theory.sinp.msu.ru} \\
{\it Skobeltsyn Institute of Nuclear Physics,
Lomonosov Moscow State University, Moscow 119991, Russia}
\end{small}
\end{center}

\vspace{0.3 cm}

\begin{footnotesize}
\noindent
{\bf Abstract.} In this work, first of all, we analyze a number of hidden aspects of the concept of particle oscillations. The key element of this concept, which do not comply with the principle of least action, is the notion of a mixture of particles, introduced by Gell-Mann and Pais for neutral $K$-mesons. It has been proven that the law of conservation of energy-momentum in the processes of electron neutrino production does not allow solving the problem of solar neutrinos based on the assumption of Gribov and Pontecorvo about their oscillations. It has been established that the consequences of Wolfenstein's equation contradict the results of the SNO and Super-Kamiokande collaborations and that the assertion by Mikheev and Smirnov on the conversion of solar neutrinos is erroneous. Another part of the work is devoted to a logically clear solution to this problem based on the hypothesis of the existence of a new interaction, the carrier of which is a massless pseudoscalar boson, which has a Yukawa coupling with electron neutrinos and nucleons. At each act of interaction of an electron neutrino with the nucleons of the Sun, caused by such interaction, the handedness of the neutrino changes from left to right and vice versa, and also the neutrino energy decreases. The hypothesis provides good agreement between the theoretical and experimental values of the rates of all five observed processes with solar neutrinos. This serves as a significant criterion for both the confidence of such a solution to the problem of solar neutrinos and the confidence of the existence of a new interaction.  
\end{footnotesize}

\vspace{0.5 cm}

\begin{small}

\begin{center}
{\large \bf 1. Introduction}   
\end{center}

   The emergence of the solar neutrino problem is caused by the report \cite{1} about the absence of observed transitions of $^{37}$Cl into $^{37}$Ar under the action of such neutrinos. It have been stated the upper limit for such a transition rate as 3 SNU (1 SNU is 10$^{-36}$ captures per target atom per second), while the rate predicted by Bahcall \cite{2} was 30$^{+30}$$_{-15}$ SNU.

   The reaction to this report was well conveyed by Reines \cite{3}: "It is interesting to note that if a positive result were obtained in the Davis's experiment, we would suddenly face the problem of whether it is due to the Sun or something else. A negative result is important because that it allows us to assert, in the event of a further positive result in a neutrino experiment, that the effect is caused precisely by solar neutrinos.This is significant, because the Universe is full of surprises ... When in February 1972 we discussed this issue at a conference in Irvine, it was interesting to observe how, in search of the reasons for the discrepancy, astrophysicists pointed to specialists in the nucleus, the latter to neutrino physicists, and those in turn to astrophysicists". These words were said by Reines during the discussion of Pontecorvo's report at the seminar on the $\mu - e$ problem. It is noteworthy that Reines does not say a word about the solar neutrinos oscillations, the assumption of which was put forward three years earlier \cite{4}. 

   It is noteworthy that none of the participants at the Irvine conference expressed a sufficiently natural assumption about the existence of some new interaction involving solar neutrinos. Herewith it would be enough to take into account Davis's result \cite{1} to establish the form of such interaction.

   As for Pontecorvo's assumption about solar neutrino oscillations, according to which the initial electron neutrino during its motion turns into a superposition of both electron and muon neutrinos, it inherits the concept of Pais and Piccioni about the oscillations of neutral $K$-mesons \cite{5}. Pontecorvo's assumption gained increasing popularity among the scientific community only after the first results of the Super-Kamiokande collaboration \cite{6}, which confirmed Davis's reports. There is widespread that the problem of solar neutrinos has been solved on the basis of the concept about their oscillations.

   But at the same time, among the hundreds of publications devoted to this concept, there is not a single one where the its results for the rates of each of the five observed processes with solar neutrinos at optimal values of free parameters in comparison with experimental results were demonstrated. The absence of such numbers deprives us of logical criteria that the problem of solar neutrinos has really been solved based on the assumption about their oscillations.

   We now claim that Pontecorvo's assumption was doomed to failure from the very beginning, since it inherited the erroneous Gell-Mann-Pais assertion that there existed, along with true particles, a "mixture of particles" \cite{7}. It seems useful to note both the groundlessness of Gell-Mann-Pais's assertion about the conservation of isospin in processes involving new particles, and the complete disregard by Gell-Mann-Pais of the principle known as Occam's razor.

\begin{center}
{\large \bf 2. Neutral $K$ mesons}
\end{center}

   Great thinkers, including Aristotle, Occam, Newton, Leibniz, considered the minimum of entities involved in explaining phenomena to be the most important criterion for the truth of assertions about them. Thus, Isaac Newton wrote: {\it We are to admit no more causes of natural things than such as are both true and sufficient to explain their appearances. To this purpose the philosophers say that Nature does nothing in vain, and more is in vain when less will serve; for Nature is pleased with simplicity, and affects not the pomp of superfluous causes} (I. Newton. The mathematical principles of natural philosophy. Book III. Rule I.) The assertion of sufficient reason acquired a rigorous mathematical form due to the introduction by Leibniz in 1669 of the notion of action as a functional of the quantities characterizing the process, and due to the formulation of the principle of least action by Montpertuis in 1744. Then the action was expressed through the Lagrangian, and the requirement of its minimality led to the Euler equation describing the transformation of the physical process.

   Since that time, all successful constructions in fundamental physics have been based on the principle of least action. Nevertheless, the notion of a mixture of particles introduced by Gell-Mann and Pais \cite{7} and the concept of oscillations of mixtures (hereinafter simply called particle oscillations), proposed by Pais and Piccioni \cite{5}, which obviously do not comply with this principle, has received wide support in the scientific community over time.

   The production of neutral hyperons and neutral $K$-mesons in collisions of $\pi^{-}$-mesons with protons in a cloud chamber, which has been observed since 1951, aroused various theoretical discussions for a long time. They especially dealt with the question of conservation of the isotopic spin at the $\pi^{-} p$-interaction vertex with the production of the hyperon $\Lambda^{0}$ and the neutral $K$-meson decaying into $\pi^{+} \pi^{-}$. 

   At that time, the notion of isospin has been applied only to nucleons and $\pi$-mesons. As for new hadrons, it would be correct first of all to find out the possibility of assigning a certain spin and its third projection to one or another hadron. The need for such finding out was not discussed by anyone at that time. Real neutral mesons $K ^{0}_{S}$ and $K^{0}_{L}$, as we now know, cannot be assigned a specific third projection of isospin. The solution to the question of isospin conservation in processes involving new hadrons had to be attributed to future research.

   Gell-Mann and Pais, not having a sufficient reason, had believed that isospin is conserved at the $\pi^{-} p$-interaction vertex, and that the $K$-meson, produced at it and denoted below as $\Theta$ (instead of the original $\theta$), has a third isospin projection $I_{3}$ equal to  -1/2. In order to smooth out the discrepancy between this $I_{3}$ value of the produced $K$-meson and the zero $I_{3}$ value of the $\pi^{+} \pi^{-}$ system into which it decays, Gell-Mann and Pais had believed that, after its production, the $K$ meson transforms into a superposition of the meson $\Theta$ and its antimeson $\bar{\Theta}$ and that, at the moment of decay $t_{0}$, it becomes a meson $\Theta_{1}$
\begin{equation}
\Theta^{0}_{1}(t_{0}) = (\Theta^{0}(t_{0}) + \bar{\Theta}^{0}(t_{0}))/\sqrt{2}.
\label{1}
\end{equation}
Orthogonal superposition gives meson $\Theta_{2}$,
\begin{equation}
\Theta^{0}_{2}(t_{0}) = (\Theta^{0}(t_{0}) - \bar{\Theta}^{0}(t_{0}))/\sqrt{2} i,
\label{2}
\end{equation}
having its own decay modes.

   Thus, the theoretical prescriptions in \cite{7} led to an excess of neutral $K$ mesons.

   Since each of the quanta $\Theta^{0}_{1}$ and $\Theta^{0}_{2}$ can be assigned its own lifetime and mass, Gell-Mann and Pais consider them to be true "particles" and treat $\Theta^{0}$ and $\bar{\Theta}^{0}$ as "mixtures of particles".

   The time transformation of the wave function of a particle with mass $m$ is well known: its initial value acquires an additional phase factor $\exp(-iE(t-t_{0}))$, where $E=\sqrt{m^{2}+ p^{2}}$ and $p$ is momentum modulus. The solution to the question of the time dependence of the wave function of a "mixture of particles" was proposed by Pais and Piccioni \cite{1}. Using the relations (\ref{1}) and (\ref{2}) give the following time transformation of $\Theta^{0}$:
$$
\Theta^{0}(t_{0}) = (\Theta^{0}_{1}(t_{0}) + i\Theta^{0}_{2}(t_{0}))/ \sqrt{2} \rightarrow (\Theta^{0}_{1}\exp(-iE_{1}(t-t_{0})) + i\Theta^{0}_{2}\exp( -iE_{2}(t-t_{0})))/\sqrt{2} = $$
\begin{equation}
= ((\Theta^{0}(t_{0}) + \bar{\Theta}^{0}(t_{0}))\exp(-iE_{1}(t-t_{0})) + (\Theta^{0}(t_{0}) - \bar{\Theta}^{0}(t_{0}))\exp(-iE_{2}(t-t_{0}))) /2.
\label{3}
\end{equation}

   Hence, under the condition $p_{1}=p_{2}\equiv p$, $E_{1} \gg m_{1}$ and $E_{2} \gg m_{2}$, the probability of detecting a meson $\Theta^{0}$ at time $t$ is an oscillating quantity:
\begin{equation}
P(\Theta^{0}, t) = \cos^{2}\frac{(m_{1}^{2}-m_{2}^{2})}{4p}(t-t_{ 0}).
\label{4}
\end{equation}

   {\it The opinion of Gell-Mann and Pais about the conservation of isospin at the $\pi^{-} p$-interaction vertex, being erroneous, required the introduction of a fictitious (false, non-existing in nature) neutral $K$-meson. This entailed false consequences: the need to introduce an excess of neutral mesons, to introduce of the division of particles into true particles and their "mixtures", respectively, having and not having certain masses, and to introduce of the concept of oscillation of "mixtures of particles", that is not described by any Lagrangian or Euler equations.}

\begin{center}
{\large \bf 3. Gribov-Pontecorvo's assumption about neutrino oscillations}
\end{center}

   Gribov and Pontecorvo suggested that the problem of solar neutrinos can be solvable based on the assumption about the existence of neutrino oscillations \cite{4}. It arose under the impression of the assertions of Gell-Mann, Pais and Piccioni about neutral $K$-mesons. Two new neutrinos $\nu_{1}$ and $\nu_{2}$ with masses $m_{1}$ and $m_{2}$ were introduced in addition to the family of known neutrinos $\nu_{e}$ and $ \nu_{\mu}$. The states $\nu_{1}$ and $\nu_{2}$ are given a status true particles, and the neutrinos $\nu_{e}$ and $\nu_{\mu}$ are given a status their mixtures, so that
\begin{equation}
\nu_{1} = \cos \theta \cdot \nu_{e} - \sin \theta \cdot \nu_{\mu}, \qquad
\nu_{2} = \sin \theta \cdot \nu_{e} + \cos \theta \cdot \nu_{\mu}.
\label{5}
\end{equation}
In relations (\ref{5}), the mixing angle $\theta$ is a free parameter.

   The probability of detecting neutrinos $\nu_{e}$ with momentum modulus $p$ at time $t$ is an oscillating quantity:
\begin{equation}
P(\nu_{e}, t) = 1 - \sin^{2} 2\theta \sin^{2}\frac{(m_{1}^{2}-m_{2}^{2}) }{4p}t.
\label{6}
\end{equation}

   The most important element in our analysis of the Gribov-Pontecorvo concept of neutrino oscillations is to find out the feasibility of the law of conservation of energy-momentum in the processes of production of electronic neutrino as a mixture of two new neutrinos with different masses. To do this, it is enough to consider the process of neutron decay 
\begin{equation}
n \rightarrow p + e^{-} + \bar{\nu}_{e},
\label{7}
\end{equation}
supposing that the antineutrino in it is a mixture of antineutrinos $\bar{\nu_{1}}$ and $\bar{\nu_{2}}$ with masses $m_{1} $ and $m_{2}$. For such a process to be admissible, it is necessary to satisfy the energy-momentum conservation law for each component of the electron antineutrino, $\bar{\nu_{1}}$ and $\bar{\nu_{2}}$. Due to the difference in their masses, this would obviously be impossible in the case of well-defined masses of all particles from the process (\ref{7}). However, the neutron is unstable, and in accordance with the Heisenberg uncertainty, its mass values have a Gaussian distribution with a width $\Gamma$ related to its lifetime $\tau$ by the equality $\Gamma \tau = 1$. Since the lifetime of a neutron is \cite{8} $\tau_{n} = 879$ s, its width has the value $\Gamma = 7.5 \cdot 10^{-19}$ eV. In what follows, we will provide the momenta moduli {p} and masses {m} of particles with a subscript reflecting their name.

   The fulfillment of the law of conservation of energy in the process (\ref{7}) would be possible if, in the Gaussian distribution of the neutron mass, there were two such masses $m_{n1}$ and $m_{n2}$ that the equalities
\begin{equation}
m_{n1}=\sqrt{p_{p}^{2}+m_{p}^{2}}+\sqrt{p_{e}^{2}+m_{e}^{2}}+\sqrt{p_{\nu}^{2}+m_{1}^{2}}
\label{8}
\end{equation}
and
\begin{equation}
m_{n2}=\sqrt{p_{p}^{2}+m_{p}^{2}}+\sqrt{p_{e}^{2}+m_{e}^{2}}+\sqrt{p_{\nu}^{2}+m_{2}^{2}},
\label{9}
\end{equation}
related to the neutron rest system, are satisfied. Since neutrinos capable of manifesting themselves in experiments have momentum moduli significantly greater than the masses $m_{1}$ and $m_{2}$, then from the equalities (\ref{8}) and (\ref{9}) we have
\begin{equation}
m_{n1}-m_{n2}= (m_{1}^{2}-m_{2}^{2})/2p_{\nu}.
\label{10}
\end{equation}

   The momentum modulus $p_{\nu}$ has the greatest value in the process (\ref{7}) when the electron momentum modulus is zero. Then using the law of conservation of momentum-energy gives 
\begin{equation}
(p_{\nu})_{\rm max}=(m_{n}-m_{e})/2-m_{p}^{2}/2(m_{n}-m_{e})=0.782 {\rm MeV}. 
\label{11}
\end{equation}
We will suppose that the largest value of the mass difference $|m_{n1}-m_{n2}|$ is equal to the width of the neutron $\Gamma$. From here and from the equalities (\ref{10}) and (\ref{11}), we obtain the following inequality
\begin{equation}
|m_{1}^{2}-m_{2}^{2}| < \Gamma \cdot (p_{\nu})_{\rm max}= 5.9 \cdot 10^{-13} {\rm eV}^{2}.
\label{12}
\end{equation}
Due to the identity of any two electron neutrinos, the restriction (\ref{12}) on the masses of the components of such neutrinos do not depend on the process in which $\nu_{e}$ is produced.
 
   Let us note here, the statement of the SNO \cite{9} and Super-Kamiokande \cite{10} collaborations that the quantity $|m_{1}^{2}-m_{2}^{2}|$ has a value of the order of $10^{-5}$. The law of conservation of energy-momentum, discussed above, prohibits an electron neutrino from being a mixture of new neutrinos with such a difference in the squares of their masses.

   {\it So, the law of conservation of momentum energy in the processes of electron neutrino production does not allow solving the problem of solar neutrinos based on the assumption of Gribov and Pontecorvo about their oscillations.}

   Let us note the fact that the concept of neutrino oscillations may contradict the law of conservation of energy-momentum has been noted repeatedly in the literature previously, for example, in \cite{11} and \cite{12}. To avoid violation of the energy-momentum conservation law, these works proposed to consider neutrino wave packets within the framework of quantum mechanics or neutrino propagators at limited time intervals within the framework of quantum field theory, but no numerical estimates were provided.

\begin{center}
{\large \bf 4. Contradiction between the consequences of the Wolfenstein's equation and the results of experiments with solar neutrinos}
\end{center}

   Our work \cite{13} presents a numerical analysis of the consequences of the Wolfenstein's equation for solar neutrinos \cite{A} with values of the oscillation parameters given by the Super-Kamiokande collaboration \cite{10}
\begin{equation}
\Delta m^{2} = 4.8 \cdot 10^{-5} \; {\rm eV}^{2}, \qquad \sin^{2} \theta = 0.334 .
\label{13}
\end{equation}
This analysis is based on the adiabotic approximation, which assumes that over several oscillation lengths the matter density in the Sun can be considered constant. After finding the interval of oscillation lengths corresponding to all observed values of the solar neutrino energy, we are convinced of the validity of this assumption when, using the numerical values of the dependence of the density of matter on its distance from the center of the Sun \cite{14}, we find that its relative change $\Delta \rho / \rho$ during the length of one oscillation does not exceed $1.7 \cdot 10^{-3}$.

   A detailed description of the change in the probability $P_{e}(t)$ that the solar neutrino state at the time moment t is electron is given. The lengths of probability oscillations $P_{e}(t)$ lie in the range from 134 km to 1028 km. Each individual neutrino oscillation in the Sun, starting from the first one, begins and ends with a purely electronic state of the neutrino. Its ending at time $t_{n}$, due to continuity, serves as the beginning of the next oscillation , i.e. $P_{e}(t_{n}) = 1$. If the completion of the next oscillation at the moment $t_{n_{S}}$ occurs at the surface of the Sun, where the density of matter is zero, then the probability amplitude of the electronic state of the solar neutrino at the moment $t_{E}$ of reaching the experimental setup on Earth is given by the following expression
\begin{equation}
P_{e}(t_{E}) = \frac{1}{2} (1+ \cos^{2} 2\theta)+\frac{1}{2} (1- \cos^{2} 2\theta)\cos(\Delta m^{2}/2E)(t_{E}-t_{n_{S}}).
\label{14}
\end{equation}

   It takes into account the fact that according to the standard solar model (SSM), each neutrino source $s$ has a sufficiently wide spherically symmetric distribution over the solar volume \cite{14}. For neutrinos from the decays of $^{8}{\rm B}$ the distribution width at half amplitude level is 37000 km, and for neutrinos from $p-p$ collision it is 77000 km. Thus, for parallel neutrino fluxes, generated in different places of the Sun and entering an experimental setup on Earth the difference in the numbers of neutrino oscillations along the various trajectories can be several tens or hundreds. The variability of trajectories leads to the fact that the values of the cosine in Eq. (\ref{16}) cover the entire interval from -1 to 1. The electron neutrino flux at the Earths surface coming from the source $s$, $\Phi_{e}(s)$, is found by summing over the neutrino fluxes along various trajectories multiplied by the corresponding probabilities
\begin{equation}
\Phi_{e}(s)=\int P_{e}(t_{E}) d \Phi(s) = \bar{P_{e}}(t_{E}) \int d \Phi(s) = \bar{P_{e}}(t_{E}) \Phi(s),
\label{15}
\end{equation}
where $\Phi(s)$ is the neutrino flux from the source $s$ given by the standard solar model. For the average probability $\bar{P_{e}}(t_{E})$, we take the right side of Eq. (\ref{14}) averaged over the cosine argument in the range from 0 to $2\pi$.

  As a result of summation over neutrino sources distributed in the Sun we obtain survival probability $P_{ee}$ of electron components of these neutrinos at the Earth`s surface the following expression
\begin{equation}
P_{ee} = \bar{P_{e}}(t_{E}) = \frac{1}{2} (1+ \cos^{2} 2\theta) = 1- \frac{1}{ 2}\sin^{2} 2\theta
\label{16}
\end{equation}
which coincides with the expression for the average probability of detecting electronic components in neutrinos oscillating in a vacuum.

   The probability $P_{ee}$ (\ref{16}) obviously contradicts the fact that, in three of the five observed processes with solar neutrinos, the ratio $\cal{K}$ of every experimental rate to the theoretical one calculated in the framework of the SSM is no more than 0.5. So, omitting orders in the values of that or another rate and giving first a reference to the experimental work, and
then to the theoretical one, we have: for the nuclear transitions $^{37}{\rm Cl} \rightarrow ^{37}{\rm Ar}$ the ratio $\cal{K}$ is $(2.56 \pm 0.24)/(7.9 \pm 2.6) = 0.32 \pm 0.14$ \cite{15} vs \cite{14}; for the elastic scattering of solar neutrinos on electrons $\nu_{e} + e^{-} \rightarrow \nu_{e} + e^{-}$ the ratio $\cal{K}$ is $(2.32 \pm 0.07)/(5.79 \times (1 \pm 0.23)) = 0.40 \pm 0.10$
\cite{10} vs \cite{16}; for the deuteron disintegration by the charged currents $\nu_{e} + D \rightarrow e^{-} + p + p$ the ratio $\cal{K}$ is $(1.76 \pm 0.11)/(5.79 \times (1 \pm 0.23)) = 0.30 \pm 0.09$ \cite{17} vs \cite{16}.

   For the nuclear transitions $^{71}{\rm Ga} \rightarrow ^{71}{\rm Ge}$ the ratio $\cal{K}$ within the error does not contradict the value in Eq. (\ref{16}). Namely, $\cal{K}$ is $(65.4 \pm 2.9)/(131 \pm 10) = 0.50 pm 0.06$ \cite{18} vs \cite{16}. The experimental and theoretical rates of the process of deuteron disintegration by neutral currents, $\nu_{e} + D \rightarrow \nu_{e} + n + p$, are close to each other.

   Thus, the Wolfenstein's equation, as the only analytical tool available to the concept of solar neutrino oscillations, with parameters from the Super-Kamiokande collaboration and with similar parameters from the SNO collaboration, is unable to explain the process rates measured in the experiments of the same collaborations. So how did these collaborations become owners of neutrino oscillation parameters (\ref{13})? We do not find the answer to this question either in the articles of the SNO collaboration \cite{17}, \cite{19}, and \cite{9} with the results of three phases of their experiments, or in the articles of the Super-Kamiokande collaboration \cite{20}, \cite{21}, \cite{22}, and \cite{10} with the results of four stages of their experiments. These collaborations provide a comprehensive and very detailed description of the setup of their experiments and the criteria for finding the number of events caused specifically by solar neutrinos. {\it It is extremely surprising that, while interpreting the discrepancies between the theoretical and experimental values of event rates as a result of neutrino oscillations, Super-Kamiokande and SNO collaborations do not give any indication of analytical formulas or procedures connecting the experimental rates of events in each of the implemented experiments with the parameters of solar neutrino oscillations, but are limited to only references to works \cite{A} and \cite{23} that do not contain such formulas or procedures}.

\begin{center}
{\large \bf 5. Assertion by Mikheev and Smirnov on the conversion of solar neutrinos}
\end{center}

   At the end of the 1970s, the problem of solar neutrinos acquired a clear numerical outline, consisting in the fact that the experimentally measured rate of the transitions of chlorine into argon under the action of solar neutrinos is approximately 1/3 of the theoretical rate found in the framework of the standard solar model. Namely, Davis announced \cite{24} that the measured rate of these transitions is $2.2 \pm 0.3$ SNU, and Bahcall et al reported \cite{25} that the calculated rate is $7.5 \pm 1.5$ SNU. At the same time, at any values of the oscillation parameters in their standard sense, the probability that the initial solar electron neutrino appears electronic at the place of its registration periodically takes on the value 1 and, therefore, when averaging over the time of one oscillation, it is equal no less than 1/2.

   Under these conditions, an extraordinary scenario of neutrino oscillations was announced by Mikheev and Smirnov in \cite{23}. Its main assertion is that the initial beam of electron neutrinos, after passing through a certain layer of matter, is almost completely transformed into a beam of muon neutrinos. The only analytical tool used in the work \cite{23} is the Wolfenstein's equation in the adiabatic approximation, the validity of which was confirmed by us in the vicinity of the oscillation parameters given in (\ref{13}).

   Here we will pay attention to the fact that using only one variant of the value of the mixing angle $\theta_{m}$ of massive neutrinos in the medium $\nu_{1m}$ and $\nu_{2m}$ in situation, where there are two variants of its value, plays a key role in declaring the metamorphosis of the solar neutrino.

   On solar neutrino trajectories containing a point with extreme values of the oscillation length and the quantity $\sin^{2} 2\theta_{m}$, the last quantity first increases with increasing matter density from $\sin 2\theta$ to 1, and then decreases to a certain value $c_{0}$, which is the smaller, the greater the density of matter in the place of neutrino production. The first option for changing the angle $2\theta_{m}$ is as follows: with increasing density of matter, it increases from $2\theta$ to $\pi/2$, and then decreases to the value $\arcsin c_{0}$. The second option is this: the angle $2\theta_{m}$ with increasing density of matter invariably increases from $2\theta$ to $\pi - \arcsin c_{0}$.

   In \cite{23}, an option of changing the angle $\theta_{m}$ from $\theta$ to $\pi/2 - (\arcsin c_{0})/2$ is chosen without mentioning the existence of the first variant.  Now comes the culminating moment of metamorphoses, described fragmentarily both by Mikheev and Smirnov in the article \cite{23}, and 34 years later by Smirnov in a recent report \cite{26}, which I present in an orderly and complete form. For a small value of $c_{0}$, expansion of the purely electronic state of a solar neutrino into states of neutrinos $\nu_{1}$ and $\nu_{2}$ in a medium is written as
$$
\nu_{e} = \cos [\pi/2-(\arcsin c_{0})/2] \cdot \nu_{1m}(\rho_{\rm max}) +
\sin [\pi/2-(\arcsin c_{0})/2] \cdot \nu_{2m}(\rho_{\rm max}) \approx $$
\begin{equation}
\approx [c_{0}/2]\cdot \nu_{1m}(\rho_{\rm max})+
\nu_{2m}(\rho_{\rm max})
\label{17}
\end{equation}
The transformation of the state of a neutrino as it moves in the Sun is determined by changes in the mixing angle $\theta_{m}$ and changes in the states of massive neutrinos. At the moment when the solar neutrino is near the exit from the Sun, its state $\nu_{\rm sol}$ is given by the formula
\begin{equation}
\nu_{sol} = \cos \theta \cdot \nu_{1m}(\rho=0) + \sin \theta \cdot \nu_{2m}(\rho=0).
\label{18}
\end{equation}
Since the decomposition of the electron neutrino (\ref{17}) is completely dominated by the neutrino $\nu_{2}$, and the decomposition (\ref{18}) at small values of $\theta$ is dominated by the neutrino $\nu_{1}$, Mikheev and Smirnov conclude, that the final neutrino is a muon neutrino. 

   The significant difference in the values of the coefficients related to the same massive neutrinos in the expansions (\ref{17}) and (\ref{18}), enhanced in \cite{23} by the smallness of the angle $\theta$ adopted there, is entirely due to the choosing the second option of the dependence of the mixing angle $\theta_{m}$ on the density of the medium.

   In the first option of the dependence of the mixing angle $\theta_{m}$ on the density of the medium, instead of the relation (\ref{17}), we have the following equality
\begin{equation}
\nu_{e} = \cos [(\arcsin c_{0})/2] \cdot \nu_{1m}(\rho_{\rm max}) +
\sin [(\arcsin c_{0})/2] \cdot \nu_{2m}(\rho_{\rm max}) \approx \nu_{1m}(\rho_{\rm max})+
[c_{0}/2]\cdot \nu_{2m}(\rho_{\rm max})
\label{19}
\end{equation}

   Now in both expansions (\ref{19}) and (\ref{18}), the same neutrino $\nu_{1}$ dominates and, following Mikheev and Smirnov, one should say that both the initial and final neutrinos are electronic. The conclusion that the final neutrino is a muon neutrino vanishes like a dream.
Moreover, the probability of detecting a neutrino in the $\nu_{e}$ state at a certain moment in time can, due to oscillations, be either greater or less than the probability of detecting a neutrino in the $\nu_{\mu}$ state at the same moment in time, therefore the change in the ratio of these probabilities cannot be a conversion criterion.

   {\it The assertion of Mikheev and Smirnov about the transformation of an electron neutrino into a muon neutrino when passing through a medium turns out to be wrong}.

   The transition from the concept of solar neutrino oscillations, which had at least a minimal number of analytical formulas, to the picture with conversion deprives the game with solar neutrinos of any analytical support. Thus, Smirnov together with Krastev write in the article \cite{27}: "However the specific mechanism of the conversion has not yet been identified".

\begin{center}
{\large \bf 6. Hypothesis about the existence of a new interaction involving electron neutrino}
\end{center}
 
   It would seem that the whole history of physics pointed to the high probability that some new for us, rather hidden, interaction is responsible for the emergence of the solar neutrino problem. Surprisingly, that the question of new interaction was not even raised for decades.

   The hypothesis about the existence of an interaction, which is carried by a massless pseudoscalar boson having Yukawa couplings with an electron neutrino, proton, and neutron (with u- and d-quarks), described by the following relativistically invariant Lagrangian
\begin{equation}
 L = ig_{\nu_{e}ps}\bar{\nu}_{e}\gamma^{5}\nu_{e}\varphi_{ps}+
ig_{Nps}\bar{p}\gamma^{5}p\varphi_{ps}-ig_{Nps}\bar{n}\gamma^{5}n\varphi_{ps},
\label{28}
\end{equation}
and not coupled with the electron at the tree level, was published in the complete form in work \cite{31}.

   The article \cite{31} contains all the essential episodes of solving the problem of solar neutrinos, based on logically clear methods of classical quantum field theory.

   Let us first note the main aspects of the separate elements of the Lagrangian (\ref{28}).

\begin{center}
{\large \bf 7. Electron neutrino as an element of the new interaction}
\end{center}

   It seems natural to believe that the electron neutrino and the electron have the same group-theoretical properties. We consider that the state of the electron neutrino is described by the bispinor representation of the Lorentz eigengroup, and its field obeys the Dirac equation. It follows that all solutions with positive energy of the massless free Dirac equation, of which two (left-handed and right-handed) can be taken for basic ones, describe various states of the same neutrino. If there is external pseudoscalar field interacting with the neutrino, then both the left and right spinors of neutrino wave vector will have nonzero values.

   The indicated equality of the electron neutrino and the electron underlies the initially $P$-invariant gauge model (corrected left-right symmetric model) of the electroweak interaction \cite{32}.

   Nevertheless, apparently, the dominant opinion is about the group-theoretic inequality of the electron neutrino and the electron and about the absence of a right-handed massless neutrino. The formation of such an opinion was influenced by two circumstances. Firstly, there is an abundance of publications in which the neutrino appears as a Majorana fermion, despite the impossibility of assigning a Lagrangian to it and despite the lack of experimental evidence of double neutrinoless beta decay of nuclei, which is allowed for the Majorana nature of the neutrino. Ignoring the principle of least action has no experimental support! Secondly, according to the established $V-A$ structure of the weak interaction Lagrangian, beta decays of nuclei produce either left-handed neutrinos or right-handed antineutrinos, and such decays are the sources of significant fluxes of solar neutrinos and reactor antineutrinos.

   In the framework of the theory with the Lagrangian (\ref{28}), the right-handed state of a neutrino is formed from the left-handed state of the same neutrino when it emits a real or virtual massless pseudoscalar boson $\varphi_{ps}$, which is due to the Lorentzian structure of the pseudoscalar current:   
\begin{equation}
\bar{\psi} (p_{2}) \gamma^{5} \psi (p_{1}) = \bar{\psi}_{R} (p_{2}) \gamma^{5} \psi_{L} (p_{1}) - \bar{\psi}_{L} (p_{2}) \gamma^{5} \psi_{R} (p_{1}).
\label{29}
\end{equation} 

   It is this picture that is realized when neutrinos move inside the Sun. At every collision a neutrino with nucleons, caused by the exchange of a massless pseudoscalar boson, its handedness changes from left to right and vice versa. As a result, the fluxes of left- and right-handed electron neutrinos near the Earth's surface are approximately equal. A more accurate estimate carried out in \cite {33} gives the ratio of these fluxes a value of 0.516:0.484.

   The contribution from right-handed solar neutrinos to the charged current processes and to the elastic scattering on electrons is extremely small, since such neutrinos are not coupled with intermediate bosons of the standard model. They can be only coupled with very heavy intermediate bosons of the initially $P$-invariant (left-right symmetric) model $W_{R}$ and $Z_{LR}$. The analysis of the nucleosynthesis in the early Universe \cite{34} and the electroweak fit \cite{8} give correspondingly the following estimate: $M_{W_{R}} > 3.3$ TeV and $M_{Z_{LR}} > 1.2$ TeV. It is noteworthy that the right handednesses of solar neutrinos nevertheless manifest themselves, namely in the process of the deuteron disintegration by the neutral currents due to the exchange of massless pseudoscalar boson, where it contributes at approximately the same level as the left  handedness of solar neutrinos.

\begin{center}
{\large \bf 8. Massless pseudoscalar boson as an element of the new interaction}
\end{center}

   Initially assigning zero mass to the pseudoscalar boson guarantees its stability.

   Let us note first of all that at large distances $r$ from the nucleon, its potential energy in a pseudoscalar field decreases as $r^{-3}$ \cite{35}, while it has the behavior $r^{ -1}$ in a long-range field.

   A massless pseudoscalar boson cannot be considered as some realization of the Peccei-Quinn axion \cite{36}. Peccei and Quinn, without any reason, expressed the opinion that the QCD Lagrangian may contain a term that violates $CP$-invariance. To avoid this fictitious violation, they postulated the existence of a pseudo-Goldstone boson called an axion. The axion must decay into two gamma quanta. It is quite natural that numerous experiments have not revealed any traces of such a particle, since the axion hypothesis violates the principle of sufficient reason, which was discussed at the beginning of section 2. This assumption belongs to the same category of judgments that ignore fundamental logical rules, such as the introduction of the notion of mixtures of particles and the imposition to neutrinos of Majorana nature.

   The theoretical rate of observed processes with solar neutrinos depends on the product of the Yukawa's coupling constants of a massless pseudoscalar boson with an electron neutrino and with nucleons. Our calculations in \cite{31} give the following estimate for it:
\begin{equation}
\frac{g_{\nu_{e}ps}g_{Nps}}{4\pi} = (3.2 \pm 0.2) \cdot 10^{-5}.
\label{30}
\end{equation}
Each of the constants $g_{\nu_{e}ps}$ and $g_{Nps}$ individually can a priori have values in a fairly large range. Meanwhile, processes in which the massless pseudoscalar boson $\varphi_{ps}$ interacts only with nucleons or only with electron neutrinos present undoubted interest both for the physics of the Sun and for the physics of the Earth.

   First of all, we note the fact that, at collisions of nucleons inside the Sun, not only gamma quanta, but also massless pseudoscalar bosons $\varphi_{ps}$ produce. The kinematics of both processes are the same, therefore the spectra of bosons $\varphi_{ps}$ are close to the spectra of gamma quanta at their production, and the intensity of bosons is related to the intensity of gamma quanta at the place of their production as the square of the Yukawa coupling constant of the boson $\varphi_{ps }$ with nucleons $g_{Nps}^{2}/4\pi$ to the electromagnetic interaction constant $\alpha$. When moving in matter, the interaction of massless pseudoscalar bosons with nucleons leads to their conversion into gamma quanta: $\varphi_{ps} + N \rightarrow \gamma + N$. This circumstance in itself, apparently, affects the difference between the spectrum of bosons $\varphi_{ps}$ and the spectrum of gamma quanta at the exit from the Sun. In any case, one can expect a significant flux of free massless pseudoscalar bosons near the Earth's surface. However, their registration as independent particles is hardly possible due to the aforementioned conversion into gamma quanta, which, strictly speaking, is determined by the total cross section of such conversion.

\begin{center}
{\large \bf 9. Kinematics of elastic scattering of a solar neutrinos on nucleons}
\end{center}

   The change in the handedness of an electron neutrino at every act of its collision with nucleons, caused by interaction (\ref{28}), is the most significant factor in reducing the rate of observed processes with solar neutrinos compared to those calculated within the SSM.

   The second factor in reducing the rate of observed processes is the decrease in neutrino energy at its elastic scattering on a resting nucleon with mass $M$. If the incident neutrino has energy
$\omega_{1}$, then the energy of the scattered neutrino $\omega_{2}$, regardless of its handedness, can take evenly distributed value in interval
\begin{equation}
\frac{\omega_{1}}{1+2\omega_{1}/M} \leq \omega_{2} \leq \omega_{1}.
\label{31}
\end{equation}
As the neutrino energy decreases, the cross sections for the observed processes also decrease, withal in different ways for different processes.

   Average value of the relative change in neutrino energy, as a result of one collision with a nucleon,
\begin{equation}
\frac{\Delta \omega_{1}}{\omega_{1}} = \frac{\omega_{1}}{M}\cdot \frac{1}{1+2\omega_{1}/M}
\label{32}
\end{equation}
is for solar neutrinos from $^{8}{\rm B}$ (their average energy equals 6.7 MeV) one order of magnitude higher than that for neutrinos from $p-p$ (their maximum energy equals 0.423 MeV) and from $^ {7}{\rm Be}$ (with energy 0.384 or 0.862 MeV). Neutrinos from $^{8}{\rm B}$ play a main role in the $^{37}{\rm Cl} \rightarrow ^{37}{\rm Ar}$ transitions and reducing their energy leads to a significant decrease their rate. At the same time, neutrinos from $p-p$ and from $^{7}{\rm Be}$ give a dominant contribution to the $^{71}{\rm Ga} \rightarrow ^{71}{\rm Ge} $ transitions, and therefore their rate decreases slightly compared to that given by the SSM, what is observed in the experiment.

   So, the interaction (\ref{28}), at a qualitative level, well reproduces the degree of difference between the predicted and experimental values of the rates of various observed processes with solar neutrinos. This gives hope for the success of the discussed approach at the quantitative level.

\begin{center}
{\large \bf 10. On an approximate description of the consequences of the Brownian motion of neutrinos in the Sun}
\end{center}

   Due to collisions with nucleons caused by the interaction (\ref{28}), solar neutrinos, being inside the Sun, experience Brownian movement. A mathematically accurate description of the consequence of this movement, which consists in finding the distribution of the number of collisions of neutrinos with nucleons before leaving the Sun depending on the product of the Yukawa's constants $\beta \equiv g_{\nu_{e}ps}g_{Nps}/4\pi$ and on the neutrino energy, seems almost impossible. With that, it is clear that the success of solving the problem of solar neutrinos based on the hypothesis of the existence of interaction (\ref{28}) depends on the simplicity of one or another approximate description of the consequences of the Brownian motion of neutrinos in the Sun.

   It turns out that the required approximation is partly formed by the internal properties of the interaction (\ref{28}) together with the limited energy of solar neutrinos. Namely, the total cross section for elastic scattering of a neutrino by a nucleon at rest is given by the following expression
\begin{equation}
\sigma = \frac{(g_{\nu_{e}ps}g_{Nps})^{2}}{16\pi M^{2}} \frac{1}{1+2\omega/M} ].
\label{33}
\end{equation}
The cross section (\ref{33}) can be considered practically independent of the solar neutrino energy, since it is limited to 18.8 MeV \cite{14}. In this approximation, we can assume that the distribution of the number of collisions of neutrinos with solar nucleons does not depend on the energy of solar neutrinos. We convey the essence of the distribution and its dependence on the unknown product of Yukawa constants $\beta$ in two methods.

   The first method concerns the choice of a free parameter that replaces the product of Yukawa's constants $\beta$. We presented it in the article \cite {31}, where the role of a free parameter is played by the effective number of collisions of neutrinos with nucleons of the Sun $n_{0}$, which is assumed to be the same for all observed processes with solar neutrinos. 

   The second method is discussed in article \cite{33}. In it, the geometric distribution, well known in mathematics and reflecting intuitive expectations, protrude as a test distribution 
\begin{equation}
P_{\beta}(n) = p(1-p)^{n}, \qquad n=0,1,2, \ldots,
\label{34}
\end{equation}
The only free parameter in it, $p$, corresponds to the probability of neutrinos leaving the Sun without collisions with its nucleons.

   Let us first look at some details of the first method. As mentioned above, after one act of the elastic scattering of the neutrino on a rest nucleon, the initial fixed value of its energy is transformed into the evenly distributed energy interval (\ref{31}). After the second act of scattering, each energy value from this interval is transformed into its own interval of type (\ref{31}). Etc.

   Regarding the methods acceptable for calculations (say, in FORTRAN, as we did it), we
have considered two variants to describe the energy distribution of neutrinos, having fixed
initial energy $\omega$, after $n_{0}$ collisions with nucleons. 

   In the first variant, the energy attributed to a neutrino after each collision is equal to the mean value of the kinematic interval (\ref{31}), so that we have sequentially for zero, one, ..., $n_{0}$ collisions
\begin{equation}
\omega_{0} = \omega, \quad \omega_{1} = \omega_{0}\frac{1 + \omega_{0}/M}{1 + 2\omega_{0}/M}, \quad \ldots \quad \omega_{n_{0}} = \omega_{n_{0}-1}\frac{1 + \omega_{n_{0}-1}/M}{1 + 2\omega_{n_{0}-1}/M}.
\label{35}
\end{equation}
In this variant, the final state of the neutrino is characterized by a single energy value given
by the last term of the sequence (\ref{35}).

   In the second variant, it is assumed that, as a result of each collision with a nucleon, the
neutrino energy takes one of the two boundary values of the interval (\ref{31}) with equal probability. Due to that, after $n_{0}$ collisions the initial level of energy $\omega$ turns into a set of $n_{0} + 1$ binomially distributed (with the success probability 1/2) values which are listed below:
\begin{equation}
\omega_{1} = \omega, \omega_{2} = \frac{\omega_{1}}{1+2\omega_{1}/M} \quad \ldots \quad 
\omega_{n_{0}+1} = \frac{\omega_{n_{0}}}{1+2\omega_{n_{0}}/M}.
\label{36}
\end{equation}

   Both variants yield close results. Thus, the replacement of the energy interval (\ref{31}) by three, four, etc. equiprobable values is inexpedient. We use everywhere only the second variant,
which is more comprehensible in its logical plan than the first one.

   Calculating the rate of the observed process with solar neutrinos that had energy $\omega$ at production and experienced $n_{0}$ effective collisions with solar nucleons means, by definition, finding the sum of the rates of the process under consideration caused by neutrinos with $n_{0} + 1 $ energy levels listed in the formula (\ref{36}). The number $n_{0}$ and the numbers in the relations (\ref{36}) have nothing to do with neutrino handedness.

\begin{center}
{\large \bf 11. On finding the numerical values of the rates of observed processes}
\end{center}

   Finding the numerical values of the rates of all five observed processes requires
at least 17 separate operations. Let us limit ourselves to the example of the $^{37}{\rm Cl} \rightarrow ^{37}{\rm Ar}$ transition process, having the threshold energy 0.814 MeV. This process is caused by neutrinos from 6 sources: $^{8}{\rm B}$, $^{7}{\rm Be}$, $^{15}{\rm O}$, $^{13}{\rm N}$, $pep$, and $hep$. we select only the first one.

   The energy values of the neutrino from $^{8}{\rm B}$, extending from 0 to about 16 MeV, are given in the table of Ref. \cite{37} as the set $\omega_{i}^{B} = i\Delta^{B}$, where $i = 1, \ldots, 160$, $\Delta B$ = 0.1 MeV. Their distribution is expressed through probability
$p(\omega_{i}^{B})$ of that neutrinos possess energy in an interval $(\omega_{i}^{B}-\Delta B/2, \omega_{i}^{B}+ \Delta B/2)$. The solar neutrino flux at the Earth surface $\Phi(^{8}{\rm B})$ from the decay of $^{8}{\rm B}$ is taken equal to the central value of $5.79 \times 10^{6} (1 \pm 0.23)$ cm$^{-2}{\rm s }^{-1}$) \cite{16}.

   We use the dependence of the cross-section of the process of neutrino absorption by chlorine
on the neutrino energy $\sigma^{\rm Cl}(\omega)$, presented in the table IX and partly in the table VII of Ref. \cite{14}. We assign for this cross-section a linear interpolation in each energy interval. 

   Within the framework of the method with the effective number of collisions of neutrinos with solar nucleons, the formula for calculating the rate of transitions $^{37}{\rm Cl} \rightarrow ^{37}{\rm Ar}$ caused by neutrinos from $^{ 8}{\rm B}$ has the following form
\begin{equation}
V (^{37}{\rm Cl} | {\rm B}) = k\Phi(^{8}{\rm B}) \sum_{i=1}^{160} \Delta^{B} p(\omega_{i}^{B})
\sum_{n=1}^{n_{0}+1} \frac{n_{0}!}{2^{n_{0}}(n-1)!(n_{0}+1-n)!}
\sigma^{\rm Cl}(\omega_{n,i}^{B}),
\label{37}
\end{equation}
where energy values $\omega_{n,i}^{B}$ is the member of the sequence (\ref{36}) with number
$n$, and in the first term of the sequence the quantity $\omega$ needs to be set equal to $\omega_{i}^{B}$. The coefficient $k$ in (\ref{37}) is equal to 0.5 in the case of equality of fluxes of left- and right-handed neutrinos at the Earth's surface, as it assume in the work \cite{31}, or 0.516 in the case of accepting the results of the work \cite{33}.

   Within the framework of the method with a geometric distribution the formula for the rate of transitions $^{37}{\rm Cl} \rightarrow ^{37}{\rm Ar}$ contains only those collisions after which the neutrino has left handedness:
\begin{equation}
V({}^{37}{\rm Cl} \ | \ {\rm B}) = \sum_{k=0}^{\infty} p(1-p)^{2k} 
\Phi({}^{8}{\rm B})\sum_{i=1}^{160}\Delta^{B}p(\omega_{i}^{B})
\sum_{n=1}^{2k+1}\frac{(2k)!}{2^{2k}(n-1)!(2k+1-n)!}
\sigma^{\rm Cl}(\omega_{n,i}^{B}), 
\label{38}
\end{equation}

   The results of theoretical calculations of the rates of all observed processes with solar neutrinos, based on the Lagrangian (\ref{28}) and on the distribution (\ref{34}), are in the best agreement with the experimental results if $p = 0.062$.

   Formulas similar to relations (\ref{37}) and (\ref{38}) are valid for all sources and for all observed processes, except for deuteron disintegration caused by the exchange of a massless pseudoscalar boson $\varphi_{ps}$. Since the deuteron binding energy is 2.225 MeV, this process can only be caused by neutrinos from $^{8}{\rm B}$ and from $hep$. It consists of two non-interfering subprocesses. One of these subprocesses is caused by the $Z$-boson exchange that occurs between a left-handed solar neutrino and a deuteron. The second subprocess is caused by the exchange of a massless pseudoscalar boson $\varphi_{ps}$, occurring between a deuteron and a solar neutrino with any handedness. The cross section of this subprocess depending on the neutrino energy, $\sigma^{\rm nc(ps)}$, present in the work \cite{31} by an analytical formula and the tabulated form. The contribution to the rate of neutrino deuteron disintegration of this subprocess, caused by neutrinos from $^{8}{\rm B}$, is given by the following expression
\begin{equation}
V({\rm D} \ | \ {\rm B}) = \sum_{k=0}^{\infty} p(1-p)^{k} 
\Phi({}^{8}{\rm B})\sum_{i=1}^{160}\Delta^{B}p(\omega_{i}^{B})
\sum_{n=1}^{k+1}\frac{(k)!}{2^{k}(n-1)!(k+1-n)!}
\sigma^{\rm nc(ps)}(\omega_{n,i}^{B}), 
\label{39}
\end{equation}

The corresponding neutrino contribution from $hep$ can be neglected.

\begin{center}
{\large \bf 12. Comparison of theoretical and experimental numbers for the rates of all five observed processes}
\end{center}

   We find it appropriate to finally present theoretical and experimental numbers for the rates of all observed processes taken from our work \cite{33}.

\begin{center}
\begin{tabular}{lccccccc}
\multicolumn{8}{c}{{\bf Table 1.} The rate of transitions ${}^{37}{\rm Cl} \rightarrow {}^{37}{\rm Ar}$ in SNU.} \\
\hline
\multicolumn{1}{l}{} 
&\multicolumn{1}{c}{${}^{8}{\rm B}$} 
&\multicolumn{1}{c}{${}^{7}{\rm Be}$}
&\multicolumn{1}{c}{${}^{15}{\rm O}$}
&\multicolumn{1}{c}{$pep$}
&\multicolumn{1}{c}{${}^{13}{\rm N}$}
&\multicolumn{1}{c}{$hep$}
&\multicolumn{1}{c}{Total} \\ 
\hline
Experiment \cite{15} &  &  &  &  &  &  & $2.56 \pm 0.16 \pm 0.16$ \\
Eq. (\ref{37}), $k=0.5$, $n_{0} = 11$ & 1.97 & 0.43 & 0.17 & 0.11 & 0.04 & 0.01 & 2.72 \\
Eq. (\ref{37}), $k=0.516$, $n_{0} = 12$ & 1.95 & 0.43 & 0.17 & 0.11 & 0.05 & 0.01 & 2.72 \\
Eq. (\ref{38}), $p=0.062$ & 2.02 & 0.42 & 0.17 & 0.11 & 0.04 & 0.01 & 2.77 \\
\hline
\end{tabular}
\end{center}

\begin{center}
\begin{tabular}{lcccccccc}
\multicolumn{9}{c}{{\bf Table 2.} The rate of transitions ${}^{71}{\rm Ga} \rightarrow {}^{71}{\rm Ge}$ in SNU.} \\ 
\hline
\multicolumn{1}{l}{}
&\multicolumn{1}{c}{$p$-$p$}
&\multicolumn{1}{c}{${}^{7}{\rm Be}$} 
&\multicolumn{1}{c}{${}^{8}{\rm B}$} 
&\multicolumn{1}{c}{${}^{15}{\rm O}$}
&\multicolumn{1}{c}{${}^{13}{\rm N}$}
&\multicolumn{1}{c}{$pep$}
&\multicolumn{1}{c}{$hep$}
&\multicolumn{1}{c}{Total} \\ 
\hline
Experiment \cite{38} &  &  &  &  &  &  &  & $62.9^{+6.0}_{-5.9}$ \\
Experiment \cite{18} &  &  &  &  &  &  &  & $65.4^{+3.1}_{-3.0}{}^{+2.6}_{-2.8}$ \\
Eq. (\ref{37}), $k=0.5$, $n_{0} = 11$ & 34.6 & 17.2 & 4.9 & 2.8 & 1.7 & 1.4 & 0.02 & 62.6 \\
Eq. (\ref{37}), $k=0.516$, $n_{0} = 12$ & 35.7 & 17.7 & 4.9 & 2.9 & 1.7 & 1.4 & 0.02 & 64.4 \\
Eq. (\ref{38}), $p=0.062$ & 35.6 & 17.6 & 5.0 & 2.8 & 1.7 & 1.4 & 0.02 & 64.2 \\
\hline
\end{tabular}
\end{center}

\begin{center}
{{\bf Table 3.} Effective fluxes of neutrinos $\Phi_{eff}^{\nu e}({}^{8}{\rm B})$ found from the process} \\
\begin{tabular}{lccccc}
\multicolumn{6}{l}{$\nu_{e} e^{-}\rightarrow \nu_{e} e^{-}$ ($E_{c}$ is given in MeV, and the fluxes are in units of $10^{6}$ ${\rm cm}^{-2}{\rm s}^{-1}$).} \\  
\hline
\multicolumn{1}{l}{References}
&\multicolumn{1}{c}{$E_{c}$} 
&\multicolumn{1}{c}{Experimental}
&\multicolumn{1}{c}{Eq. (\ref{37}), $k=0.5,$} 
&\multicolumn{1}{c}{Eq. (\ref{37}), $k=0.516,$}
&\multicolumn{1}{c}{Eq. (\ref{38}),} \\
&& results & {$n_{0} = 11$} & {$n_{0} = 12$} & {$p=0.062$} \\
\hline
SK III \cite{22} & 5.0 & $2.32\pm 0.04 \pm 0.05$ & 2.27 & 2.29 & 2.26 \\
SNO I \cite{17} & 5.5 &$2.39^{+0.24}_{-0.23}{}^{+0.12}_{-0.12}$ & 2.19 & 2.20 & 2.17\\
SNO II \cite{39} & 6.0 &$2.35\pm 0.22\pm 0.15$ & 2.10 & 2.10 & 2.09\\
SNO III \cite{9} & 6.5 &$1.77^{+0.24}_{-0.21}{}^{+0.09}_{-0.10}$ & 2.01 & 2.01 & 2.01\\
\hline
\end{tabular}
\end{center}

\begin{center}
{{\bf Table 4.} Effective fluxes of neutrinos $\Phi_{eff}^{cc}({}^{8}{\rm B})$ found from the process} \\
\begin{tabular}{lccccc}
\multicolumn{6}{l}{$\nu_{e}D \rightarrow  e^{-}pp$ ($E_{c}$ is given in MeV,and the fluxes are in units of $10^{6}$ ${\rm cm}^{-2}{\rm s}^{-1}$).} \\ 
\hline
\multicolumn{1}{l}{References}
&\multicolumn{1}{c}{$E_{c}$} 
&\multicolumn{1}{c}{Experimental}
&\multicolumn{1}{c}{Eq. (\ref{37}), $k=0.5,$} 
&\multicolumn{1}{c}{Eq. (\ref{37}), $k=0.516,$}
&\multicolumn{1}{c}{Eq. (\ref{38}),} \\
&& results & {$n_{0} = 11$} & {$n_{0} = 12$} & {$p=0.062$} \\
\hline
SNO I \cite{17} & 5.5 & $1.76^{+0.06}_{-0.05}{}^{+0.09}_{-0.09}$ & 1.86 & 1.85 & 1.88 \\
SNO II \cite{39} & 6.0 & $1.68^{+0.06}_{-0.06}{}^{+0.08}_{-0.09}$ & 1.77 & 1.74 & 1.80 \\
SNO III \cite{9} & 6.5 & $1.67^{+0.05}_{-0.04}{}^{+0.07}_{-0.08}$ & 1.66 & 1.62 & 1.72 \\
\hline
\end{tabular}
\end{center}

\begin{center}
{{\bf Table 5.} Effective fluxes of neutrinos $\Phi_{eff}^{nc}({}^{8}{\rm B})$ found from the process} \\
\begin{tabular}{lccccc}
\multicolumn{6}{c}{$\nu_{e}D \rightarrow  \nu_{e}np$ (the fluxes are in units of $10^{6}$ ${\rm cm}^{-2}{\rm s}^{-1}$).} \\
\hline
\multicolumn{1}{l}{}
&\multicolumn{1}{c}{Exchange}
&\multicolumn{1}{c}{Exchange}
&\multicolumn{1}{c}{Exchange} 
&\multicolumn{1}{c}{Exchange}
&\multicolumn{1}{c}{Sum} \\
& by $Z$ & by $Z$ & by $Z$ & by $\varphi$ & \\
& Eq. (\ref{37}), $k=0.5$ & Eq. (\ref{37}), $k=0.516$& Eq. (\ref{38})& Eq. (\ref{37}), (\ref{39})& \\
\hline
SNO I \cite{17} & & & & & $5.09^{+0.44}_{-0.43}{}^{+0.46}_{-0.43}$ \\
SNO II \cite{39} & & & & & $4.94^{+0.21}_{-0.21}{}^{+0.38}_{-0.34}$  \\
SNO III \cite{9} & & & & & $5.54^{+0.33}_{-0.31}{}^{+0.36}_{-0.34}$ \\
\hline
$n_{0} = 11$ & 2.10 & & & 2.87 & 4.98 \\
$n_{0} = 12$ & & 2.11 & & 2.85 & 4.96 \\
Eq. (\ref{38}), (\ref{39}) & & & 2.13 & 2.80 & 4.92 \\
\hline
\end{tabular}
\end{center}

   As can be seen from the above tables, the theoretical rates of all five observed processes with solar neutrinos, obtained in both methods of describing the consequences of the Brownian motion of neutrinos in the Sun, are in good agreement both with each other and with the experimental rates. At that, apparently, a finer agreement is inherent in the method with an effective number of collisions equal to 12. The agreement between the two methods indicates two facts. Firstly, the assumption of the same value of the effective number of collisions for all observed processes successfully reflects reality, since in the geometric distribution method there is no dependence on the type of the observed process. Second, the geometric distribution is a acceptable approximation of the true distribution.

\begin{center}
{\large \bf 13. Conclusion}
\end{center}

   The problem of solar neutrinos discovered by Davis allows for one of two fundamentally different solutions.

   One solution announces the disappearance of some electron neutrinos on their path from the production place in the Sun to the installation on Earth due to their transitions into muon or tau neutrinos. The neutrino oscillation assumption, put forward by Gribov and Pontecorvo, considers electron neutrinos as mixtures of two new massive neutrinos. The notion of a mixture of particles, introduced earlier by Gell-Mann and Pais, contradicts the principles of classical logic. The Gribov-Pontecorvo assumption, despite hundreds of publications devoted to it, is not confirmed by numbers reflecting comparison of theoretical and experimental results for each of the five observed processes with solar neutrinos.

   Another solution considers that the flux of solar electron neutrinos is unchanged throughout their entire path, but such characteristics of their states as handedness and energy undergo changes. At the exit from the Sun there are fluxes of both left- and right-handed electron neutrinos with reduced energy. This solution is based on the hypothesis about the existence of a new interaction involving electron neutrinos and nucleons. In its implementation, logically clear methods of classical quantum field theory are used. The obtained good agreement between theoretical and experimental numbers for all five observed processes. This allows us to hope that the solution to the solar neutrino problem has acquired a complete form.

\end{small}
\end{document}